% 14-09-20
\magnification=1200 
\input epsf.tex  %macro per inserire le figure
%---------------------
\def\Enu{{\rm E}_{\nu}}
\def\Enup{{\rm E}_{\nu+1}}
\def\Enum{{\rm E}_{\nu-1}}
%%%%%%%%%%%%%%%%%%%%

\centerline{\bf A FAMILY OF EXPONENTIAL INTEGRALS}\par
\centerline{\bf  SUGGESTED BY STELLAR DYNAMICS}\par 
\vskip 0.5truecm 
\centerline{Luca Ciotti}\par
\centerline{luca.ciotti@unibo.it}\par
\vskip 0.5truecm 
\centerline{Department of Physics and Astronomy,
University of Bologna}\par 
\centerline{via Gobetti 93/3, I-40129 Bologna, Italy}\par
\centerline{(September 14, 2020)} 
\vskip 1truecm 
\centerline{\bf Abstract}
\vskip 0.5 truecm

While investigating the generalization of the Chandrasekhar (1943)
dynamical friction to the case of field stars with a power-law mass
spectrum and equipartition Maxwell-Boltzmann velocity distribution, a
pair of 2-dimensional integrals involving the Error function occurred,
with closed form solution in terms of Exponential Integrals (Ciotti
2010). Here we show that both the integrals are very special cases of
the family of (real) functions
$$
I(\lambda,\mu,\nu; z) :=\int_0^zx^{\lambda}\,\Enu(x^{\mu})\,dx=
{\gamma\left({1+\lambda\over\mu},z^{\mu}\right) +
z^{1+\lambda}\Enu(z^{\mu})\over 1+\lambda + \mu (\nu -1)},
\quad \mu>0,\quad z\geq 0,
\eqno (1) 
$$
where $\Enu$ is the Exponential Integral, $\gamma$ is the incomplete
Euler gamma function, and for existence $\lambda >\max \left\{-1,-1-
\mu(\nu -1)\right\}$.  Only in one of the consulted tables a related
integral appears, that with some work can be reduced to eq.~(1), while
computer algebra systems seem to be able to evaluate the integral in
closed (and more complicated) form only provided numerical values for
some of the parameters are assigned. Here we show how eq.~(1) can in
fact be established by elementary methods.

\vskip 1.truecm 
\centerline{\bf 1. Introduction}
\vskip 0.5 truecm 

Two interesting integrals, that can be expressed in closed form in
terms of the Error Function and of the Exponential Integral, were
encountered while generalizing the Chandrasekhar (1943) dynamical
friction formula to the case of a test mass moving in a field of stars
with a power-law mass spectrum, and equipartition Maxwell-Boltzmann
velocity distribution (eqs.~[30]-[31] in Ciotti 2010). They both
belong to the family of functions in eq.~(1): quite surprisingly, this
simple-looking identity is not found in the most important tables of
integrals (e.g., Erd\'elyi et al. 1953, Gradshteyn and Ryzhik 2007,
Prudnikov et al. 1990), and neither the latest releases of Mathematica
and Maple seem to be able to recover the general result, but only
particular cases for numerical values of some of the parameters. In
the following I show how the identity in eq.~(1) can be established
with elementary methods.

\vskip 1.5truecm  
\centerline{\bf 2. Some preliminary material}
\vskip 0.5 truecm  
 
For succesive use, we report the relevant identities obeyed by the
Exponential Integrals. They are defined for $\Re(z) >0$ as
$$
\Enu(z) := \int_1^{\infty}
t^{-\nu}e^{-tz}dt=z^{\nu -1}\Gamma(1-\nu,z),
\eqno (2)
$$
(e.g., Abramowitz \& Stegun, Chapter 5; Arfken \& Weber 2005, Exercise
8.5.8; Erd\'elyi et al. 1953, Vol.2, Chapter 9; see also
https://functions.wolfram.com, https://dlmf.nist.gov).

The last expression above, where $\Gamma(1-\nu,z)$ is the incomplete
right Euler Gamma function, is obtained with an obvious change of
integration variable. The Euler incomplete left and right Gamma
functions (over the reals) can be expressed in integral form as
$$
\gamma (a,x):=\int_0^xt^{a-1}{\rm e}^{-t}dt,\quad 
\Gamma(a,x):=\int_x^{\infty}t^{a-1}{\rm e}^{-t}dt, 
\eqno (3) 
$$
where $\Re(a)>0$ for convergence of the $\gamma$
function\footnote{$^1$}{As we do not use the continuation of the
functions to the Complex plane, from now on all the quantities are
intended reals.}, therefore they obey the relations of easy proof:
$$
\gamma (a+1,x)=a\gamma(a,x)-x^a{\rm e}^{-x},\quad 
\Gamma (a+1,x)=a\Gamma(a,x)+x^a{\rm e}^{-x}. 
\eqno (4) 
$$
and
$$
\gamma(a,x)+\Gamma(a,x)=\Gamma(a)=\int_0^{\infty}t^{a-1}{\rm e}^{-t}dt
= \gamma(a,\infty)=\Gamma(a,0),
\eqno (5) 
$$
where $\Gamma(a)$ is the complete Gamma function.

About the Exponential Integrals $\Enu$, from their integral expression
in eq.~(2) it is a simple exercise to show that
$$
{\rm E}_0(z)={{\rm e}^{-z}\over z},\quad {d\Enu(z)\over dz} =
-\Enum(z).
\eqno (6)
$$
Moreover, from integration by parts of eq.~(2), by using the first and
the second function in the integrand as differential factor, for
$z\neq 0$ one obtain respectively
$$
\Enu(z) = {{\rm e}^{-z}  -z\,\Enum(z)\over \nu -1} =
{{\rm e}^{-z} -\nu\,\Enup(z)\over z},
\eqno (7)
$$
where of course $\nu\neq 1$ in the first identity. Finally, from
standard asymptotic expansion it follows that, at the leading order
for $z\to 0$,
$$
\Enu (z)\sim\cases{
\displaystyle{
{1\over\nu-1},\quad \nu>1,}\cr\cr 
\displaystyle{-\ln z,\quad \nu=1,}\cr\cr 
\displaystyle{
{\Gamma(1-\nu)\over z^{1-\nu}},\quad \nu<1,}
}
\eqno (8)
$$
and in particular it follows that the divergence of the Exponential
Integrals near the origin gets worse for decreasing $\nu\leq 1$, an
obvious consequence of eq.~(2).  The leading-order expansions in
eq.~(8) will be used in the next Section to determine the limitations
on the values of the parameters $(\lambda,\mu,\nu)$ required for
existence of the function $I(\lambda,\mu,\nu;z)$; no special
difficulties are encountered to evaluate higher order terms, and they
can be also used to check the consistency of the recursion identities
in eq.~(7) for $z\to 0$.

\vskip 1.truecm 
\centerline{\bf 3. The parameter space} 
\vskip 0.5 truecm

Before proceeding to prove eq.~(1), it is convenient to determine the
restrictions on the values of the parameters $(\lambda,\mu,\nu)$ to
assure existence of the function $I$. Equation (8) shows that we must
consider three different cases as a function of the value of $\nu$ (a
generic real number), and in fact elementary integration shows that at
the leading order for $z\to 0^+$ and $\mu >0$
$$
I(\lambda,\mu,\nu ;z)\sim \cases{
\displaystyle{{z^{\lambda+1}\over (\lambda +1)(\nu-1)},\quad\nu>1,}\cr\cr 
\displaystyle{-{\mu\over\lambda +1}z^{\lambda+1}\ln z,\quad\nu = 1,}\cr\cr 
\displaystyle{
{\Gamma (1-\nu)\over 1+\lambda +\mu(\nu -1)}z^{1+ \lambda + \mu (\nu -1) },\quad\nu <
1,}
}
\eqno (9)
$$
provided the conditions 
$$
\lambda > \cases{
-1, \quad \nu\geq 1,\cr\cr 
-1 - \mu (\nu -1), \quad \nu\leq 1}
\qquad = \qquad \max \left\{-1,-1 -\mu(\nu -1)\right\},
\eqno (10)
$$
are satisfied (see Figure 1).

We finally notice that an obvious and useful transformation
of the function $I$ in eq.~(1) can be obtained with the change
integration variable $y=x^r$ and $r>0$
$$
I(\lambda,\mu,\nu;z) = {1\over r}I\left({\lambda -r +1\over
r},{\mu\over r},\nu;z^r\right):
\eqno (11)
$$
in particular, by setting $r=\mu$ it is always possible to reduce to
the case of integration of eq.~(1) with $\Enu$ depending linearly on
the integration variable, and this case is evaluated by Mathematica.

%---------------------------------------------------------------------
\bigskip\par 
\hskip 2 truecm 
\leavevmode 
\epsfxsize=10truecm 
\epsfbox{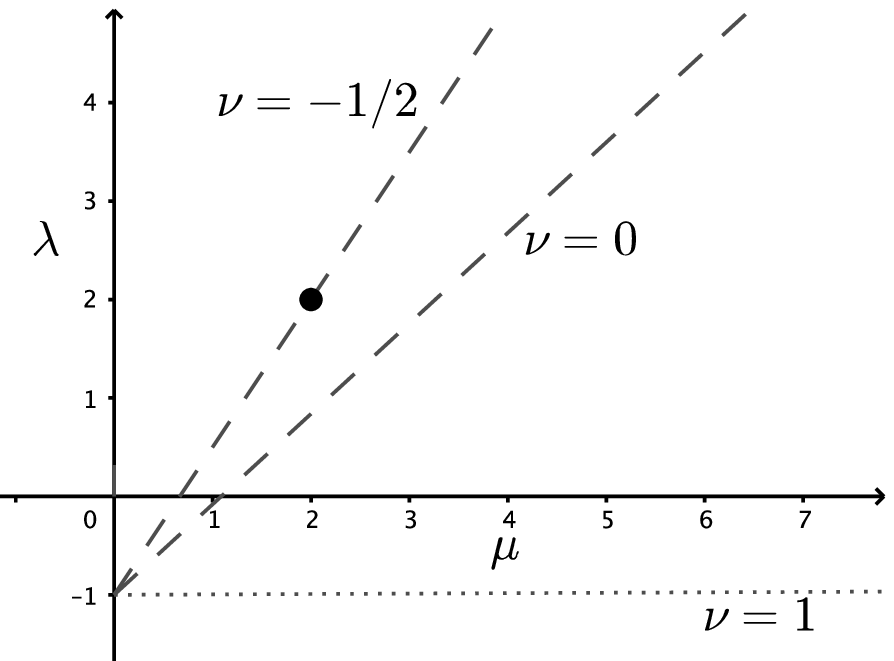}
\par 
{\bf Figure 1} The region in the $(\lambda,\mu)$ parameter space for
existence of the function $I(\lambda,\mu,\nu;z)$, as determined by
eq.~(10). For values of $\nu >1$ all points above the the horizontal
dotted line are acceptable. At decreasing $\nu$ the existence region
reduces to the points above the dashed line, here represented for $\nu
=0$ and $\nu =-1/2$. Notice that for all points above the $\nu =0$
line, also the expression in eq.~(15) can be used, and that the value
$\nu = -1/2$ is the minimum value required for existence of the
function $H$ in eqs.~(16)-(17) when $\lambda =\mu =2$ (solid dot).

\bigskip\par 
%---------------------------------------------------------------------

\vskip 0.5 truecm 
\centerline{\bf 4. A proof of identity (1)}
\vskip 0.5 truecm 

We are now in position to prove the indentity in eq.~(1) by elementary
methods.  First, as $\mu>0$, and considering that from eq.~(10)
certainly $\lambda > -1$, we can integrate by parts with $x^{\lambda}$
as differential factor, obtaining a recursion identity
$$
I(\lambda,\mu,\nu ;z) 
%{z^{\lambda+1}\Enu(z^{\mu})\over\lambda+1}
%+{\mu\over\lambda+1}\int_0^zx^{\lambda+\mu}\Enum(x^{\mu})dx 
 ={z^{\lambda+1}\Enu(z^{\mu}) + 
\mu\,I(\lambda+\mu,\mu,\nu -1;z)\over\lambda+1}. 
\eqno (12) 
$$
The first term follows from eq.~(9) and the limitations in eq.~(10),
while the second term from the second identity in eq.~(6). Then, from
the first identity in eq.~(6), and restricting (for the moment) to
$\nu\neq 1$ we have
$$
x^{\mu}\Enum(x^{\mu})={\rm e}^{-x^{\mu}}-(\nu -1)\Enu(x^{\mu}).
\eqno (13)
$$
We now multiply the identity above for $x^{\lambda}$ and integrate
over $x$, so that
$$
\mu\, I(\lambda+\mu,\mu,\nu -1;z)
=\gamma\left({1+\lambda\over\mu},z^{\mu}\right)-\mu (\nu
-1)I(\lambda,\mu,\nu,z): \eqno (14)
$$
notice that the identity also holds for $\nu =1$, so we can relax the
restriction $\nu\neq 1$. Therefore we have a second identity that can
be used with eq.~(12) to obtain the function $I(\lambda,\mu,\nu;z)$
and finally prove eq.~(1), QED.

Notice that the procedure is the same used (for example) in standard
exercises to integrate products of trigonometric functions and
exponentials.  The correctness of eq.~(1) can be verified with some
work from the second of eq. 1.2.1.1 of Prudnikov et al. (1990, Volume
2). In particular, first express the Exponential Integral in terms of
the incomplete Gamma function from eq.~(2), then change the parameters
in Prudnikov's equation as $\lambda\to\lambda +\mu (\nu -1)$,
$\alpha\to 1-\nu$, $a\to 1$, and $\nu\to \mu$, and finally combine two
Gamma functions in the incomplete $\gamma$ function from indentity
(5). Reassuringly, notice how the limitations on the parameters given
in Prudnikov, once expressed in terms of our parameters, coincide with
those given in eq.~(10).

Note that for $(1+\lambda)/\mu >1$, i.e. $\lambda > -1 +\mu$, it is
possible to apply the first recursion formula in eq.~(4) to the the
incomplete $\gamma$ function appearing in eq.~(1), and sucessively
reduce the resulting formula from the second identity in eq.~(7),
obtaining
$$ 
I(\lambda,\mu,\nu; z)=
{
{1+\lambda-\mu\over  \mu}\gamma\left({1+\lambda
-\mu\over\mu},z^{\mu}\right) -\nu z^{1+\lambda-\mu}\Enup(z^{\mu})
\over
1 +\lambda + \mu (\nu -1)}.
\eqno (15) 
$$
Of course, if $\lambda > -1 +2\mu$, the argument can be applied again
to eq.~(15), and so on, but the resulting formulae become increasingly
complicated (even if of trivial construction), and not reported here.

With the aid of eq.~(15) we can easily prove eqs.~(30)-(31) in Ciotti
(2010), that were derived by using ``ad hoc'' integration based on the
properties of the Error function.  Starting from eqs.~(16)-(21)-(29)
in Ciotti (2010), the two integrals to be solved can be written as
$$
H(y):=a c^a {4\over\sqrt{\pi}}\int_c^{\infty}r^{-\nu}dr\int_0^y t^2 
{\rm e}^{-r t^2}dt, \quad a>1,\quad c=1-{1\over a},
\eqno (16)
$$
and the two functions $H_1$ and $H_2$ of interest in Stellar Dynamics
correspond to $\nu = a -3/2$ and $\nu = a -5/2$, respectively. In the
original work the integration was performed considering first the
inner integral, and then integrating by parts over $r$ a term
involving the Error function. Here instead we invert order of
integration in eq.~(16) so that
$$
H(y)=a c^{a-\nu -1/2}
{4\over\sqrt{\pi}}
\int_0^{\sqrt{c}y}x^2\Enu(x^2)dx=
a c^{a-\nu -1/2}
{4\over\sqrt{\pi}} I(2,2,\nu;\sqrt{c}y), 
\eqno (17)
$$
where in the integral we changed variable as $x=\sqrt{c}y$. As
$\lambda = 2 > -1 +\mu =1$, it is then possible to use eq.~(15), and finally
from the identity
$$
{1\over\sqrt{\pi}}\gamma\left({1\over 2},z\right)={\rm Erf}(z),
\eqno (18)
$$
eqs.~(30)-(31) in Ciotti (2010) are recovered. Figure 1 immediately
shows (solid dot) that $a>1$ is required for existence of $H_1$, and
$a>2$ for existence of $H_2$.

\vfill\eject
%\vskip 1. truecm 
\centerline{\bf 5. Conclusions}
\vskip 0.5 truecm 

Prompted by a problem of Stellar Dynamics, an elementary derivation is
presented for the closed-form expression of a family of indefinite
integrals involving powers and Exponential Integrals. Well known
computer algebra systems seem unable to obtain the primitive in closed
form in the general case, and also for numerical values of (some) of
the parameters the resulting formulae can be quite complicated and not
easily simplified to the compact expressions in eqs.~(1)-(15), though
the numerical values are in perfect agreement. However, from the two
last identities and eqs.~(4), (7) and (11), it is expected that
general and uniform simplification procedures for the integrals
$I(\lambda,\mu,\nu;z)$ could be easily implemented in computer algebra
systems.

%\vskip 0.5truecm
%I thank ....

\vskip 0.5truecm 
\centerline{\bf 6. References}
\vskip 0.5 truecm

Abramowitz, M., and Stegun, I.A., 1970
{\it Handbook of Mathematical Functions}, Ninth Edition (Dover, New York)
\vskip 0.2truecm
Arfken, G.B., and Weber, H.J., 2005
{\it Mathematical Methods for Physicists}, Sixth Edition (Elsevier
Academic Press, Burlington, MA, USA)
\vskip 0.2truecm
Chandrasekhar, S., 1943,
{\it The Astrophysical Journal}, {\bf 97}, 263
\vskip 0.2truecm
Ciotti, L., 2010
Proceedings of the Symposium {\it Plasmas in the Laboratory and in the Universe: Interactions, Patterns, and Turbulence}, G. Bertin et al. eds, AIP Conf.Ser., vol.1242, p.117 
\vskip 0.2truecm
Erd\'elyi, A., Magnus, W., Oberhettinger, and F., Tricomi, F.G., 1953,  
{\it Higher Transcendental Functions}, (McGraw-Hill, New York)
\vskip 0.2truecm
Gradshteyn, I.S., and Ryzhik F.G., 2007,  
{\it Table of Integrals, Series, and Products - 7th Edition}, Alan Jeffrey and Daniel Zwillinger, Eds., (Elsevier, Burlington)
\vskip 0.2truecm
Prudnikov, A.P., Brychkov, Yu.A., and Marichev, O.I. 1990, 
{\it Integrals and Series}, (Gordon and Breach, New York)

\end